\documentclass[12pt]{iopart}

\usepackage{graphicx}

\begin{document}

\title[Wang X et. al.]{High-repetition-rate seeded free-electron laser with direct-amplification of an external coherent laser}

\author{Xiaofan Wang$^{1}$, Chao Feng$^{2,3}$, Bart Faatz$^{2}$, Weiqing Zhang$^{1,4}$, Zhentang Zhao$^{2,3}$}

\address{$^1$ Institute of Advanced Science Facilities, Shenzhen, Shenzhen 518000, China}
\address{$^2$ Shanghai Advanced Research Institute, Chinese Academy of Sciences, Shanghai 201210, China}
\address{$^3$ Shanghai Institute of Applied Physics, Chinese Academy of Sciences, Shanghai 201800, China}
\address{$^4$ Dalian Institute of Chemical Physics, Chinese Academy of Sciences, Dalian 116023, China}

\ead{fengchao@zjlab.org.cn, weiqingzhang@dicp.ac.cn and zhaozhentang@zjlab.org.cn}
\vspace{10pt}

\begin{indented}
\item[]February 2022
\end{indented}

\begin{abstract}
Various scientific and industrial researches such as spectroscopy and advanced nano-technologies have been demanding high flux and fully coherent EUV and X-ray radiation. These demands can be commendably satisfied with a MHz-level repetition-rate seeded free-electron laser (FEL). Dictated by the seed laser system, seeded FELs have faced obstacles for the realization of MHz repetition rate. Reducing the required peak power of an external coherent seed laser can effectively increase the repetition rate of a seeded FEL. This paper presents a novel technique that employs a long modulator as a carrier for laser amplification and electron modulation, which is quite different from nominal seeded FELs. Applications of the proposed technique into high-gain harmonic generation (HGHG) and echo-enabled harmonic generation (EEHG) are investigated. Simulation results demonstrate that seed laser power is reduced by about three orders of magnitude and the FEL radiation possesses consistently high coherence and power stability with respect to the nominal HGHG or EEHG. The proposed technique paves the way for the realization of fully coherent EUV and soft X-ray FELs with a repetition rate of MHz and an average power of about 100 W.
\end{abstract}
\maketitle
%
%
%
%
%

\section{Introduction}
MHz-level repetition-rate free-electron lasers (FELs) have been the continuous pursuit of institutions and groups around the world due to the dramatic increase in radiation pulse generation rate. Free-electron laser in Hamburg (FLASH)~\cite{honkavaara2014flash} and European X-ray free-electron laser (XFEL) are now under operation at MHz in burst-mode~\cite{altarelli2015european,decking2020mhz}, while the linac coherent light source II (LCLS-II)~\cite{galayda2014lcls}, the Shanghai high repetition rate XFEL and extreme light facility (SHINE)~\cite{zhu2017sclf} and MariX~\cite{serafini2019marix} are aiming to generate MHz FELs in continuous wave (CW) mode.

As the most common FEL operational mechanism, self-amplified spontaneous emission (SASE)~\cite{kondratenko1980generation,bonifacio1984collective} holds the potential of generating CW FELs by means of the development of superconducting technology. In a SASE FEL, lasing occurs in a single pass of a relativistic electron beam through a long periodic magnetic structure (undulator). Starting from the intrinsic electron beam shotnoise, SASE FELs have limited longitudinal coherence and large shot-to-shot power fluctuations.

Various techniques have been proposed to improve the longitudinal coherence (a bandwidth close to transform limit) of a SASE FEL. In the self-seeding scheme~\cite{feldhaus1997possible,saldin2001x,geloni2011novel}, a configuration of a monochromator sandwiched by double undulators is adopted. The SASE radiation generated by the first undulator is purified by the monochromator and then further exponentially amplified to saturation in the second undulator. Intrinsic fluctuation, however, still remains in the self-seeding scheme. Direct seeding is another way to provide a fully coherent output~\cite{lambert2008injection,togashi2011extreme}. An external coherent laser, generated from high harmonic generation (HHG) source by the means of injecting a high-power laser to a noble gas~\cite{mcpherson1987studies,ferray1988multiple,seres2005source,labaye2017extreme}, is directly injected into the undulator and amplified to saturation. However, low conversion efficiency in the X-ray wavelength of a HHG source prevent us from obtaining X-ray output of a direct seeding FEL. 

Seeded FEL schemes like high-gain harmonic generation (HGHG)~\cite{yu1991generation,yu2000high,allaria2012highly} and echo-enabled harmonic generation (EEHG)~\cite{stupakov2009using,xiang2009echo,ribivc2019coherent} have been invented to convert the external seed to shorter wavelengths. HGHG consists of two undulators separated by a chicane. An external coherent laser is used to interact with electrons in the first undulator (modulator) to generate a sinusoidal energy modulation of the electron beam at the lasing wavelength. Energy modulation is then transformed into density modulation through the dispersion element chicane. Coherent radiation at shorter wavelength is generated after the density-modulated (microbunched) electron beam traverses the second undulator (radiator), which is tuned to a high harmonic of the seed laser. HGHG ensures a high degree of longitudinal coherence but low frequency conversion efficiency. The output wavelength of HGHG can only reach the extreme ultraviolet (EUV) range~\cite{penco2020enhanced}. EEHG pushes the seeded FELs to shorter wavelength with a more complicated phase space manipulation technique. Two modulators and two chicanes are adopted in EEHG to achieve echo effect of the microbunched electron beam. The lasing of EEHG at soft X-ray wavelength regions have been achieved at the Trieste FERMI~\cite{ribivc2019coherent} and Shanghai soft X-ray free-electron laser facility~\cite{feng2019coherent}.

Highly coherent EUV and soft X-ray laser sources facilitate scientific and industrial researches such as spectroscopy and advanced nano-technologies. As one of these laser sources, HHG has been explored for applications in the nanostructure characterization and transient nanoscale dynamic of materials~\cite{kinoshita2014development,nagata2019wavelength}. The limited power of EUV radiation generated by HHG, however, hinders its application. CW seeded FELs can produce higher output at both EUV and X-ray wavelengths and are therefore more likely to be a comprehensive platform for the cutting-edge nanostructure and nanodevice researches~\cite{erik,mochi2019lensless,kudilatt2020quantum}. Other applications such as time-resolved spectroscopy~\cite{petrillo2019high}, coherent diffraction imaging~\cite{pedersoli2013mesoscale,capotondi2015multipurpose,helfenstein2017coherent} and EUV microscopy~\cite{attwood2000soft,bencivenga2015four}, would also benefit from CW seeded FELs.

Triggered by an externally coherent seed laser, the repetition-rate of a seeded FEL is dictated by the properties of the seed laser system. In the nominal HGHG or EEHG, the seed wavelength is usually at deep ultraviolet (DUV), as harmonics of commercial light sources, e.g., the Ti:sapphire laser ($\lambda\sim$800 nm) and Yb-based fiber laser ($\lambda\sim$1030 nm). Peak power of hundreds of megawatts (MW) and pulse length of dozens to hundreds of femtoseconds are preferred for seeded FELs~\cite{allaria2012highly,ribivc2019coherent,feng2019coherent}. To reach MHz-level repetition rate, tens of watts of average power is required for the DUV seed laser, which is two or three orders of magnitudes larger than currently achievable value from a commercial laser system~\cite{chang2020ultrafast}. Tremendous efforts especially the work about optical parametric chirped-pulse amplifier have been dedicated into the development of high-average-power laser system~\cite{riedel2013long,prinz2015cep,puppin2015500,hoppner2015optical,mecseki2019high}. However, it is still non-trivial for the state-of-art laser system to meet the power requirement of a seeded FEL with MHz-level repetition rate. 

Another technical route is to reform the laser-electron interaction mechanism to relax the power requirement of a high-repetition-rate seeded FEL. Several methods have been developed, such as the optical resonator~\cite{reinsch2012radiator,li2018high,petrillo2020coherent,ackermann2020novel,paraskaki2021optimization,mirian2021high}, angular dispersion enabled microbunching (ADM)~\cite{feng2017storage,wang2019angular}, self-modulation~\cite{dunning2011design,yan2021self} and optical-klystron HGHG~\cite{paraskaki2021high}. In the optical resonator, the electron beam is stored in an oscillator before it is directed to generate radiation. The repetitive laser-electron interaction is able to amplify the laser pulses and achieve considerable energy modulation on the electron beam. With combination of frequency up-conversion mechanism, e.g., HGHG or EEHG, this method could further extend the radiation to shorter wavelength. However, the repetition rate of the radiator is non-adjustable at most time. 1 MHz repetition rate indicates 300 m long resonator, and 0.5 MHz means 600 m long resonator. Also, the increased thermal issues in the cavity need to be studied to secure a stable output. ADM takes use of multi-dimensional modulation of the electron beam to relax the requirement on the laser-induced energy modulation. The need for small angular divergence of the electron beam, however, makes this method more suitable for use in storage rings. Self modulation or optical-klystron HGHG is based on pre-density modulation method. These schemes still need a laser-induced energy modulation at least half of the initial energy spread. And the correspondingly required dispersion to form the density modulation is much larger than that in nominal HGHG, which may cause the spectrum broadening induced by the energy chirps in the electron beam (Appendix).

Here, an alternative method named direct-amplification enabled harmonic generation (DEHG) is presented. DEHG is an ensemble of frequency up-conversion schemes that adopts a relative long modulator to replace the short modulator in nominal seeded FELs. This schematic simplicity allows DEHG to be compatible with nominal seed FEL schemes such as HGHG and EEHG to generate FEL output from EUV to soft X-rays. It is investigated that the long modulation process in DEHG produces a completely different physical scenario compared to the short modulation in the nominal schemes. Laser amplification and electron modulation are realized simultaneously in the modulator, and the energy-modulated electron beam together with the amplified seed laser are then guided to the downstream elements for further beam manipulation and high harmonic generation through HGHG or EEHG processes.

The requirement on the laser power is only large enough to suppress the shotnoise power (equivalent input power of SASE) of the electron beam, which is about three orders of magnitude smaller than that required in a nominal seeded FEL. Hereby, the MHz-level repetition-rate could be realized by a commercially available laser system. It is investigated that self-feedback mechanism exists in DEHG and this makes stability even better natively. Moreover, the absence of large dispersion in DEHG-based HGHG maintains the excellent longitudinal coherence of the nominal seeded FEL. In DEHG-based EEHG, the directly amplified seed laser which is reused in the second modulator inherits the current profile of the electron beam, yielding great synchronization. The proposed technique opens the way for the realization of stable MHz-level repetition-rate seeded FELs.

\section{Direct-amplification enabled harmonic generation for seeding a high-repetition-rate free-electron laser}

\begin{figure*}[!htb]
\begin{indented}
\item[]
    \includegraphics[width=1\linewidth]{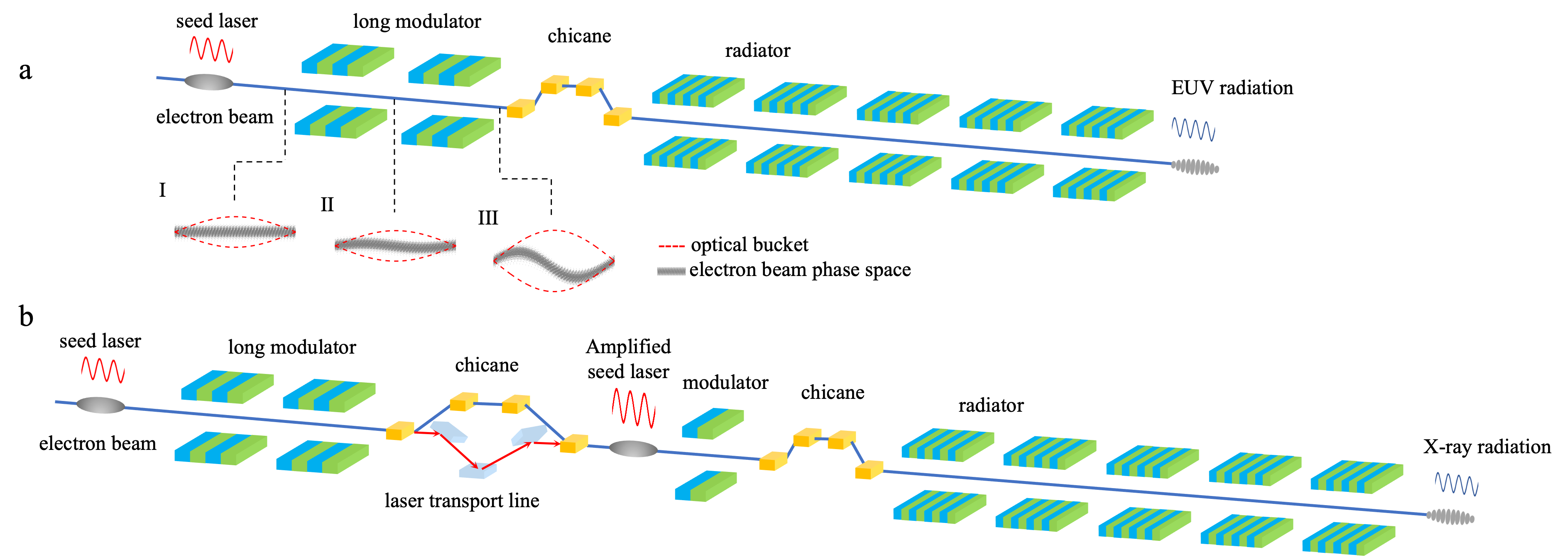}
    \centering
    \caption{Schematic layout of the direct-amplification method with its applications in HGHG (a) and EEHG (b). I-III: evolution of one optical bucket (one seed laser wavelength) and longitudinal phase space distribution of the electron beam. Optical bucket height indicates the pulse intensity of the seed laser.}
    \label{fig1}
\end{indented}
\end{figure*}

Figure \ref{fig1}
shows the schematic layout of direct-amplification method with its applications in HGHG (Fig. 1(a)) and EEHG (Fig. 1(b)). HGHG enables harmonic jump to no more than twentieth, while EEHG enlarges harmonic jump to dozens. With a seed laser at the wavelength of DUV range, HGHG with EEHG will cover the radiation wavelength from EUV to soft X-rays. For practical convenience, the undulator segment is as far as possible in accordance with the length of 4 m to customize, except one is 2.4 m. The first modulators in HGHG and EEHG consist of two undulator segments with a break length of 1 m in the middle. The break is used to put diagnostic equipment and quadrupoles. The 8-m-long modulator ensures significant amplification of the seed laser power. The second modulator in EEHG is 2.4 m. The radiators in both mechanisms consist of several 4-m-long undulator segments with a break length of 1 m in the middle. The period lengths of modulators and radiators are 8 cm and 4.1 cm, respectively. It should be mentioned that with the modulator length fixed, the laser-induced modulation can be optimized by tuning the seed laser power.

The long upstream modulator (two undulator segments) is used for seeding amplification and electron modulation, and the energy-modulated electron beam together with the amplified seed laser are then guided to the downstream elements for further beam manipulation and high harmonic generation through HGHG or EEHG processes. In the modulator, an externally coherent seed laser with a peak power larger than shotnoise is directly injected to interact with the electron beam. The long-distance modulation process can ensure sufficient energy exchange between the electron beam and the seed laser, as portrayed in Fig.~\ref{fig1}(a)(I-III). In the initial stage of modulation, the power of the seed laser grows hardly during the laser-electron interaction process while a weak energy modulation is imprinted onto the electron beam. In the latter part, the lethargy regime of seeded FEL is overcome and there is an intense energy exchange between the laser and the electron beam. At this time, the electric field increases rapidly and produces a significant sinusoidal modulation on the electron beam. Although the power of the laser enters the exponential gain regime, it is far from saturation and the rotation of the phase space is almost not perceptible, which ensures that the energy modulation of the electron beam maintains sinusoidal. This laser-electron interaction mechanism realizes the direct amplification of seed laser and effectively sinusoidal energy modulation of the electron beam.

To demonstrate the feasibility of the proposed technique, start-to-end simulations~\cite{reiche1999genesis} have been performed for both HGHG and EEHG assistant with the direct-amplification technique. HGHG will be used to produce 13.55 nm EUV radiation, while EEHG will be used to produce 6 nm soft X-ray radiation. 

\begin{figure}[!htb]
\begin{indented}
\item[]
    \includegraphics[width=1\linewidth]{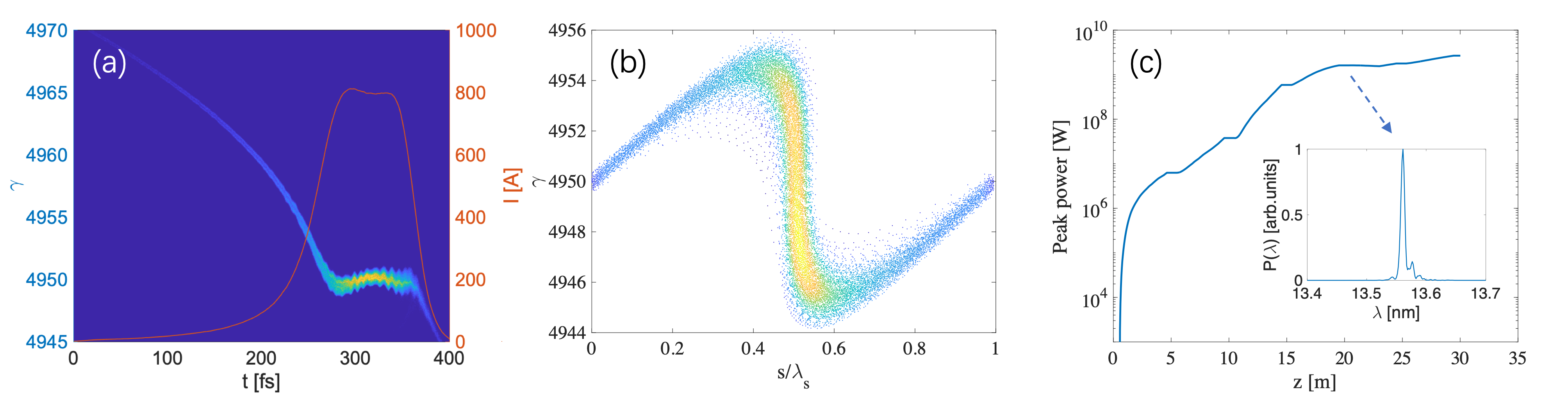}
    \caption{(a) Longitudinal phase space and current distribution of the electron beam at the entrance of the modulator ($\gamma$ is the Lorentz factor); performances of DEHG at 13.55 nm: (b) phase space distribution of the electron beam within one seed laser wavelength at the entrance of radiator (s is the longitudinal position, $\lambda_s$=257.5 nm); (c) radiation peak power along the radiator and spectrum at 20 m.}
    \label{fig2}
\end{indented}
\end{figure}

Typical parameters of a soft X-ray FEL is used here. The electron beam is at 2.5 GeV energy with spread of 0.19 MeV, duration of 110 fs (FWHM), peak current of 800 A and normalized emittance of 0.4 mm mrad. At the entrance of the modulator, the longitudinal phase space and current distribution of the electron beam are shown in Fig.~\ref{fig2}(a). The externally coherent seed laser is at 257.5 nm and is produced by the fourth harmonic generation of a commercial infrared fiber laser at 1030 nm. The pulse duration is 350 fs (FWHM). The modulator is an in-vacuum undulator with a 2.3 T peak magnetic field at 5.3 mm gap~\cite{leontowich2021lower}. K parameter of the modulator is 17.5.

For HGHG, the required harmonic number for reaching 13.55 nm is nineteenth. Owing to the large frequency up-conversion amplitude, the required energy modulation should be greater than 10 times of the energy spread. To achieve sufficient modulation of the electron beam with the 8-m-long modulator, the peak power of the seed laser is chosen to be 1 MW. Under this condition, the laser-induced energy modulation is 11.6 times of the energy spread. For a nominal HGHG whose modulator length is usually one Rayleigh length of the seed laser, simulation results show that the seed laser power needs to be larger than 640 MW to obtain the same energy modulation. Hereby the required peak power of the seed laser is reduced by almost three orders of magnitude with the proposed technique. The reduction in peak power can be traded for the increase of the repetition rate. For the direct-amplification method, the average power of the seed laser is 0.35 W under 1 MHz repetition rate. Such a power could be achieved by a commercial fiber laser system.

By setting the dispersion strength of the chicane as 34 $\mu$m, the energy-modulated electron beam becomes spatially density-modulated. Its phase space distribution is shown in Fig.~\ref{fig2}(b).
The phase space distribution is very close to that of a nominal HGHG, indicating that the direct-amplification process yields sinusoidal energy modulation. Maximum bunching factor of 3\% is achieved at 13.55 nm with the density-modulated beam. The radiator is an in-vacuum undulator with a 1.4 T peak magnetic field at 6.7 mm gap. K parameter is 5.4. In the radiator, the peak power of the EUV radiation grows exponentially and reaches saturation at 20 m with a value of 1.6 GW and pulse duration of about 60 fs (FWHM). The pulse energy is 77 $\mu$J. At this point, the spectrum bandwidth is about 1/1780 ($\Delta\lambda/\lambda$), 1.25 times to the Fourier transform limit, which manifests the great longitudinal coherence of the radiation. These results are shown in Fig.~\ref{fig2}(c). The average power of the FEL radiation at EUV reaches 77 W with 1 MHz repetition rate.

\begin{figure}[!htb]
\begin{indented}
\item[]
    \includegraphics[width=1\linewidth]{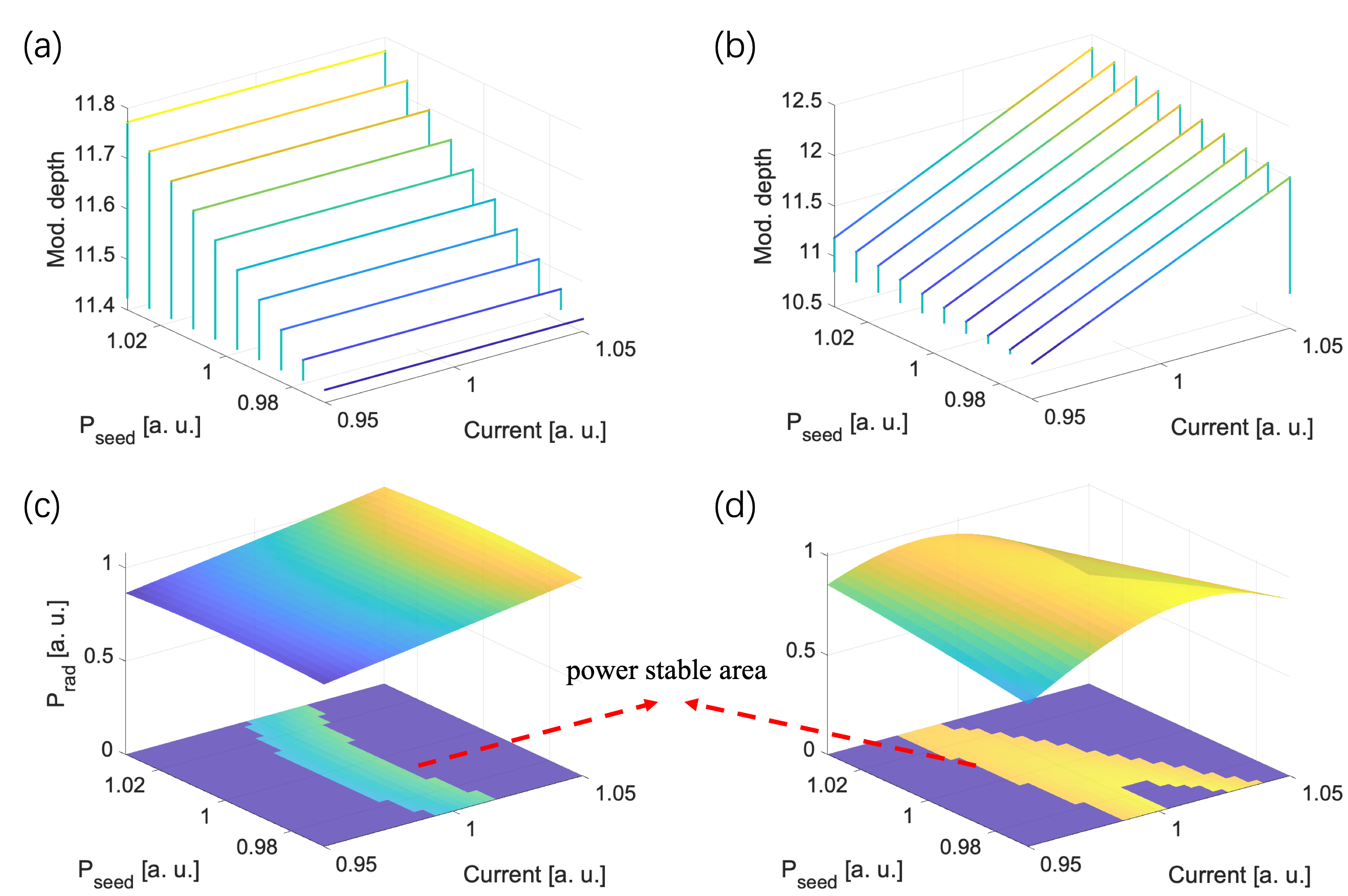}
    \caption{Comparison of modulation depth and radiation power without (a, c) and with (b, d) DEHG applied in HGHG; (a, b) modulation depth versus seed laser power and electron beam current; (c, d) radiation power versus seed laser power and electron beam current. Lower parts of (c, d) are the projections of power stable area.}
    \label{fig3}
\end{indented}
\end{figure}

Stabilities of DEHG are also studied. Figure \ref{fig3} reveals comparisons of modulation depth and radiation power without (a, c) and with (b, d) DEHG applied in HGHG.
Fig.~\ref{fig3}(a, b) present the modulation depth as functions of the seed laser power and electron beam current. For nominal HGHG, the modulation depth is related to the seed laser intensity, but not to the electron beam current. For DEHG, however, the modulation depth is correlated with both the seed laser power and the electron beam current. This phenomenon makes the radiation of DEHG more tolerant to the variation of current. The reason is that although an increase of beam current will enhance the radiation power, the coexisting increase in modulation depth and the fixed dispersion strength result in the over-compressed and a smaller bunching factor, which is not conducive to the radiation power growth. This feedback mechanism makes radiation of DEHG not naturally sensitive to the beam current. Fig.~\ref{fig3}(c, d) display the radiation power versus seed laser power and electron beam current in nominal HGHG and DEHG, respectively. One can see that in nominal HGHG, the radiated power increases as the electron beam current rises. In DEHG, however, the radiation power does not increase with current in a certain area due to the feedback mechanism. Power stable area can be regarded as a region in which the radiation power decreases by no more than 5\% due to jitters in the electron beam current and seed laser power. In this case, the power stable area of DEHG is 28\% larger than that of nominal HGHG. Further simulations reveal that this quantity becomes larger when the frequency up-conversion amplitude is lower. When the radiation frequency is at 9th of the seed laser, the power stable area of the DEHG is twice as large as that of the nominal HGHG. These results indicate the great stability of DEHG.

The above simulation results demonstrate that the proposed technique is capable of generating stable, nearly full coherent and MHz-level repetition-rate EUV radiation. Nineteenth harmonic generation is almost the limit of HGHG with the above parameters. Shorter wavelengths are no longer at the scope of a single-stage HGHG and it can be achieved through EEHG with a similar setup of the first modulator based on the direct-amplification technique, as shown in Fig.~\ref{fig1}(b). Like the nominal EEHG, two modulators, two chicanes and one radiator are arranged in the layout. The structural differences lay in the increased length of the first modulator, the added laser transport line between the two modulators, and the absence of the second external seed laser. The first modulator which consists of two undulator segments enables laser amplification and electron modulation. The amplified seed laser is then forwarded to the laser transport line to get appropriate time delay ($\sim$30 fs), in order to interact with the electron beam again in the subsequent modulator. The laser transport line mentioned here is an optical system that uses lenses and mirrors to focus and transmit the optical field with minimal power loss. With commercially available high transmittance lenses and high reflectivity mirrors, power loss can be controlled within 5\%. After the twice-modulated electron beam passes through the second chicane, a highly micro-bunched beam distribution is formed. In comparison to the nominal EEHG, whose pulse energy fluctuations largely inherit the time jitter of the second seed laser, DEHG with a naturally synchronized second seed laser is very favorable for generating stable radiation.

\begin{figure*}[!htb]
\begin{indented}
\item[]
    \includegraphics[width=1\linewidth]{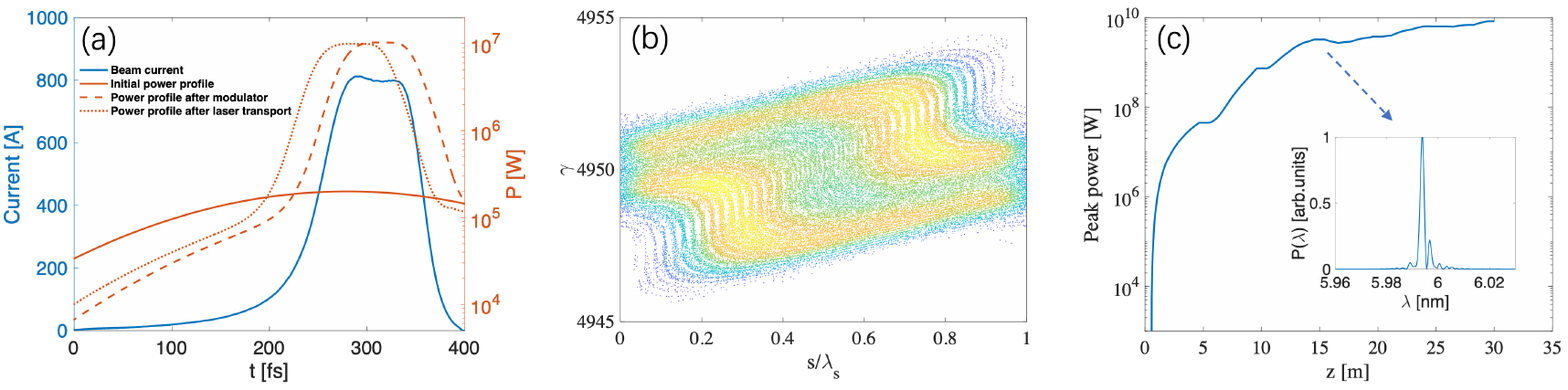}
    \caption{Performances of DEHG-based EEHG at 6 nm: (a) the beam current distribution and power profile evolution of the seed laser; (b) phase space distribution of the electron beam within one seed laser wavelength ($\lambda_s$=257.5 nm) at the entrance of radiator; (c) radiation peak power along the radiator and spectrum at 15 m.}
    \label{fig4}
\end{indented} 
\end{figure*}

Radiation at 6 nm was simulated to illustrate the capability of the proposed technique to generate soft X-ray pulses. In this case, the peak power of the seed laser is 0.2 MW, indicating average power of 70 mW under 350 fs pulse duration (FWHM) and 1 MHz repetition rate. Parameters of the electron beam and undulators are all the same with those mentioned above. After the first modulator, the laser power would be amplified to 10 MW and the electron beam would get 5.2 times of the energy spread at the same time. Amplified seed laser is then led to the laser transport line while the electron beam traverses the chicane. The power profile of the seed laser at different positions (i.e., at entrance of the 1st modulator, at exit of the 1st modulator and at entrance of the 2nd modulator) are illustrated in Fig.~\ref{fig4}(a), 
where the beam current profile is also presented as a reference coordinate. One can see clearly the enhancement of the seed laser intensity through the 1st modulator and the inheritance of the power profile to the electron beam current. The laser transport line introduces a 30 fs time delay of the laser to the electron beam without obviously decreasing the laser power. The introduced time delay can effectively counteract the slippage time in the second modulator to obtain a uniform energy modulation of the electron beam. In the second modulator, the amplified seed laser induces 2.9 times of the energy spread on the electron beam. By now, the two-stage modulation of the electron beam is accomplished by a unique external coherent laser source. For the nominal EEHG with one-Rayleigh-length modulators, however, the total required peak power of two seed lasers to obtain the same energy modulation is 170 MW, which is about three orders of magnitude higher than the value used here. 

With the dispersion strength of two chicanes as 8.1 mm and 0.18 mm, respectively, the electron beam is highly micro-bunched at 6 nm and the maximum bunching factor of 6\% is achieved. Phase space distribution of the electron beam at the entrance of radiator is presented in Fig.~\ref{fig4}(b). The distribution is very close to that of a nominal EEHG, which indicates the feasibility of the setup in Fig.~4. The radiator is an in-vacuum undulator with a 0.87 T peak magnetic field at 11 mm gap. K parameter of the radiator is 3.3. The peak power of the radiation grows exponentially in the radiator and reaches saturation at 15 m, with a saturated power of 3.2 GW and and pulse duration of about 54 fs (FWHM). The pulse energy is 0.135 mJ. At saturation, the spectrum bandwidth is about 1/3320 ($\Delta\lambda/\lambda$), 1.18 times to the Fourier transform limit. These results are shown in Fig.~\ref{fig4}(c). With a repetition rate of 1 MHz, the average power of the FEL radiation at 6 nm reaches 135 W. The radiation performances like power and spectrum bandwidth of EEHG at 6 nm are better than those of HGHG at 13.5 nm, since the bunching factor is larger and the laser-induced energy spread is smaller. It should be emphasized that self-feedback mechanism also emerges in this case, which is a hint for the generation of stable output.

\section{CONCLUSION AND OUTLOOK}

Thanks to the development of superconducting technology, MHz-level repetition-rate electron beam has come true. Seeded FELs have been proved to be convincing method to produce fully coherent EUV and soft X-ray radiation. However, limited by the average power of laser system today, seeded FEL mostly cannot run at the repetition-rate as high as the electron beam from superconducting linear accelerator. The huge challenges for reducing the required seed laser power while maintaining the great longitudinal coherence and mechanism simplicity of nominal seeded FELs has led to the invention of the proposed technique. The proposed technique, with its application either on HGHG or EEHG, reduces the requirement of the seed laser power by about three orders of magnitude and make CW seeded FELs possible for a commercially available fiber laser system. The achievable average power at EUV and soft X-ray range reaches about one hundred watts with a repetition rate of 1 MHz. The mechanical layout of the proposed technique is close to the that of nominal seeded FELs and this provides excellent conditions for subsequent proof-of-principle demonstrations and technical applications. Progress made in this paper paves the way for the development of fully coherent EUV and soft X-ray FELs with a repetition rate of MHz. 

\ack{
The authors would like to thank Guorong Wu (IASF), Wei Liu (IASF), Yong Yu (DICP), Jun Zhao (SARI), Li Zeng (IASF), Hao Sun (SINAP) and Georgia Paraskaki (DESY) for helpful discussions and useful comments related to this work. This work is supported by National Natural Science Foundation of China (11975300) and Shanghai Science and Technology Committee Rising-Star Program (20QA1410100).}


\appendix
\section*{Appendix. Radiation spectrum of an energy-chirped electron beam}
\setcounter{section}{1}

In the mechanism of high-gain harmonic generation (HGHG)~\cite{yu1991generation}, studies have shown that the radiation wavelength is very sensitive to the dispersion strength when the electron beam has an energy chirp~\cite{feng2012study,hemsing2014highly,paraskaki2019impact}. Since energy chirp of the electron beam is substantive in many FEL facilities~\cite{feng2016single,ribivc2019coherent}, the desire to reduce the wavelength shift, or spectral bandwidth, has led to the preference for low dispersion strength. In the scheme of direct-amplification enabled harmonic generation (DEHG), the required dispersion strength is almost the same as that required in nominal HGHG. In the HGHG-based pre-bunch schemes, like self-modulation~\cite{yan2021self}, optical-klystron HGHG~\cite{paraskaki2021high} or harmonic cascade~\cite{dunning2011design}, two chicanes are basically needed, and the strength of the first chicane is usually one order of magnitude larger than that required in nominal HGHG, in order to form density modulation in the fundamental or second harmonic of the seed laser. 

\begin{figure*}[!htb]
\begin{indented}
\item[]
\includegraphics[width=1\linewidth]{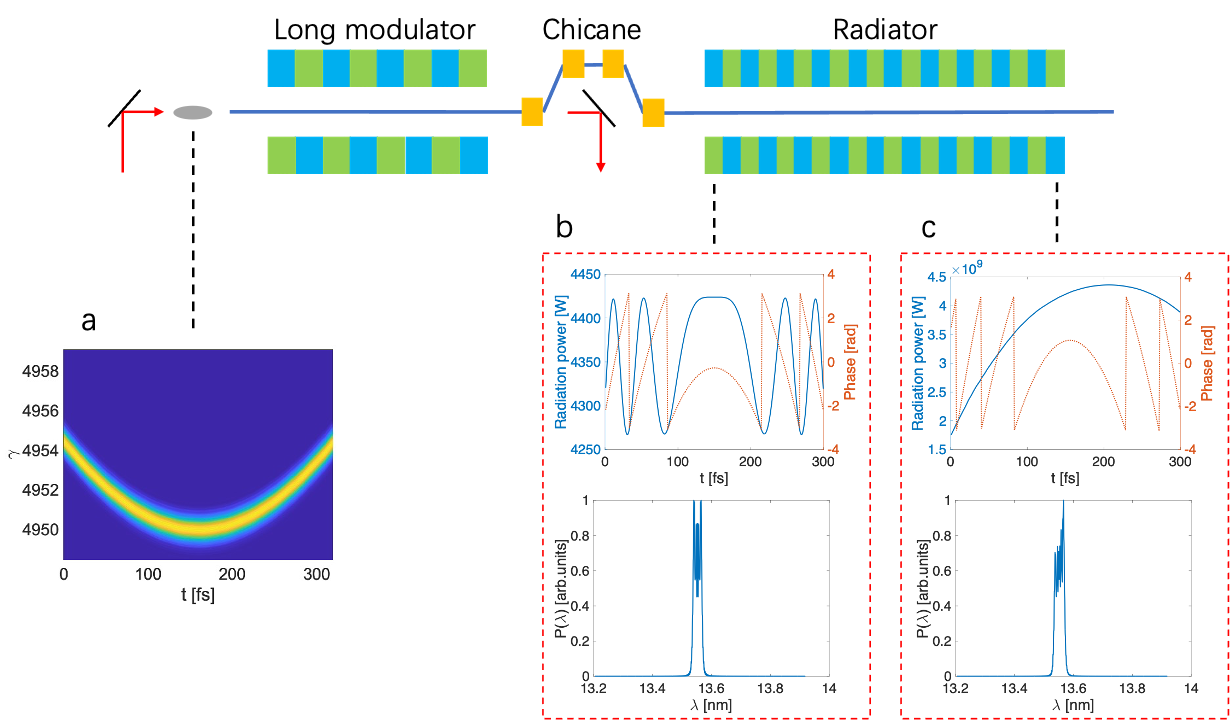}
\caption{Schematic layout of DEHG with its application in HGHG. a: The longitudinal phase space of the electron beam; In b and c, top pictures show the longitudinal profile of the radiation power (solid line) and phase (dashed line); bottom pictures show the spectrum. b: initial radiation stage; c: radiation at 20 m. Breaks between undulators are not displayed.}
\label{DEHG}
\end{indented}
\end{figure*}

\begin{figure*}[!htb]
\begin{indented}
\item[]
\includegraphics[width=1\linewidth]{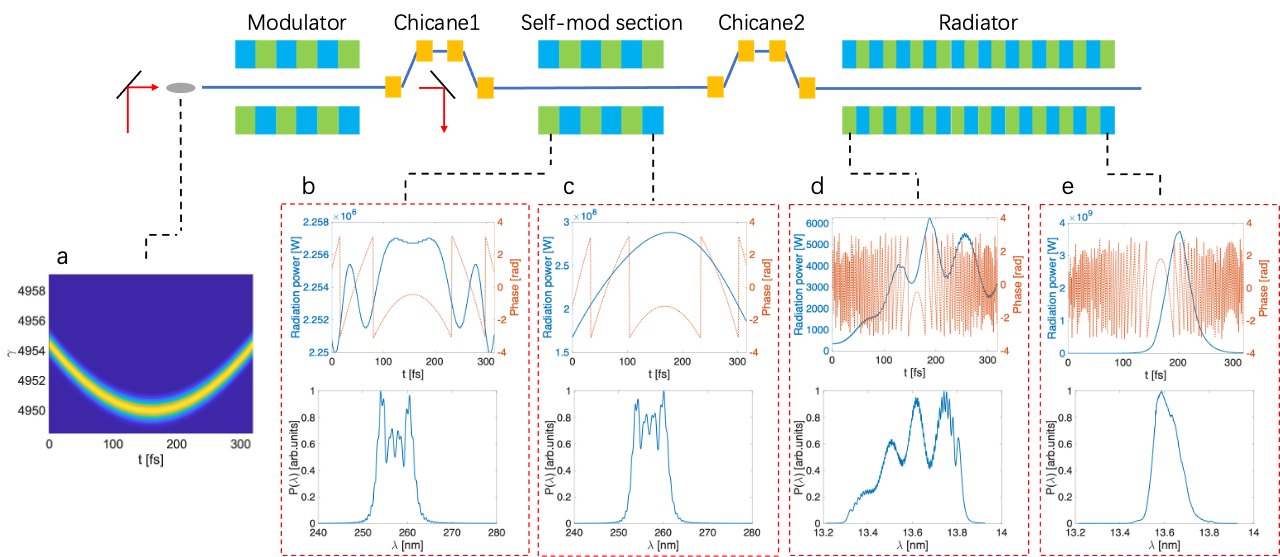}
\caption{Schematic layout of self-modulation. a: The longitudinal phase space of the electron beam; In b, c, d and e, top pictures show the longitudinal profile of the radiation power (solid line) and phase (dashed line); bottom pictures show the spectrum. Breaks between undulators are not displayed.}
\label{self_mod}
\end{indented}
\end{figure*}

In order to describe the radiation spectrum of an energy-chirped electron beam, we examine its spectrum performances obtained from DEHG followed by a comparison in self-modulation. The schematic layout of DEHG and self-modulation are illustrated in Fig.~\ref{DEHG} and Fig.~\ref{self_mod}, respectively. In DEHG, substantial energy modulation is directly achieved in the sole modulator. In self-modulation, however, it is achieved in the self-mod section and the upstream segments are acted as a pre-buncher to form microbunching in the fundamental of the seed laser. Both schemes use the same electron beam and assume it has a sinusoidal distribution in the longitudinal phase space. Electron distribution can be seen from Fig.~\ref{DEHG}(a)
or Fig.~\ref{self_mod}(a). 
The electron beam is at 2.5 GeV with a duration of 350 fs, an energy spread of 0.19 MeV and a maximum energy deviation of 2.5 MeV. For simplicity, the electron beam is assumed uniformly distributed with a current of 800 A and the seed laser is assumed much longer than the electron beam with a peak power of 1 MW at 257.5 nm. The frequency up-conversion amplitude is set to be nineteen. For the undulator part, DEHG has a modulator of 8 m and a radiator of 20 m, while self-modulation has a modulator of 1.28 m, a self-mod section (modulator) of 2.32 m and a radiator of 20 m. The period lengths of modulators and radiators are 8 cm and 4.1 cm, respectively.
 
In DEHG, the electron beam is firstly modulated by the seed laser and gets modulation about 11.6 times of its energy spread. After transported into the chicane with a strength of 34 $\mu$m, the electron beam is highly micro-bunched and the corresponding bunching factor is about 3\%. The radiation performances of the electron beam in the radiator can be seen in Fig.~\ref{DEHG}(b-c). The top pictures show the radiation power and phase, while the bottom ones show the spectrum. It can be seen that the energy chirp contributes to the local phase shift in the initial radiation, resulting in a spectrum bandwidth of 0.03 nm (FWHM). The slippage effect in the radiator may induce additional phase change during the radiation amplification process, but it can not eliminate the phase shift introduced by the energy chirp. The radiation power reaches 4.4 GW at 15 m, and the pulse energy is 1.1 mJ.

In self-modulation, the seed laser brings 0.57 times of the energy spread of the electron beam in the modulator. To convert the energy modulation to density modulation, the required dispersion strength of the first chicane is 510 $\mu$m, which is about fifteen times larger than that in DEHG. This quantity will also be the local shift in phase of the initial radiation produced in the downstream radiator. Fig.~\ref{self_mod}(b-c) shows the radiation performances of the electron beam in the self-mod section. The top pictures show the radiation power and phase, while the bottom ones show the spectrum. Note that the self-mod section resonates at the fundamental wavelength, which means the phase shift presented in Fig.~\ref{self_mod}(b-c) is 19 times larger than that in Fig.~\ref{DEHG}(b-c). The electron beam achieves energy modulation of 12.86 times of the energy spread in the self-mod section, and then traverses the second chicane with a strength of $3.3\times 10^{-5}$ m, resulting in a maximum bunching factor of 3.6\%. The radiation performances of the electron beam in the radiator is shown in Fig.~\ref{self_mod}(d-e). Compared with DEHG, the local phase shift of the radiation is much larger, indicating that the local phase is seriously affected by the large dispersion strength. The bandwidth of the spectrum at the initial radiation is 0.34 nm FWHM, ten times larger than that of DEHG. The radiation power reaches 3.8 GW at 15 m, and the pulse energy is 0.2 mJ, which is five times smaller than that of DEHG.

The phenomenon described here demonstrates the advantage of the DEHG, which effectively reduces the seed laser power without expanding the required intensity of the dispersion section with respect to nominal HGHG. Compared to the HGHG-based pre-bunch schemes, the longitudinal coherence of DEHG has a larger tolerance for energy chirp.


\section*{References}
\bibliographystyle{unsrt}
\bibliography{mybib}

\begin{thebibliography}{10}

\bibitem{honkavaara2014flash}
Katja Honkavaara, Bart Faatz, Josef Feldhaus, S~Schreiber, Rolf Treusch, and
  M~Vogt.
\newblock Flash: First soft x-ray fel operating two undulator beamlines
  simultaneously.
\newblock In {\em Proc. of FEL2014, Basel, Switzerland}, volume~14, 2014.

\bibitem{altarelli2015european}
Massimo Altarelli.
\newblock The european x-ray free-electron laser: toward an ultra-bright, high
  repetition-rate x-ray source.
\newblock {\em High Power Laser Sci. Eng.}, 3:E18, 2015.

\bibitem{decking2020mhz}
W~Decking et~al.
\newblock A mhz-repetition-rate hard x-ray free-electron laser driven by a
  superconducting linear accelerator.
\newblock {\em Nat. Photon.}, 14(6):391--397, 2020.

\bibitem{galayda2014lcls}
John Galayda et~al.
\newblock The new lcls-ii project: Status and challenges.
\newblock In {\em LINAC14, Geneva, Switzerland}, pages 404--408, 2014.

\bibitem{zhu2017sclf}
ZY~Zhu, ZT~Zhao, D~Wang, Z~Liu, RX~Li, LX~Yin, and ZH~Yang.
\newblock Sclf: An 8-gev cw scrf linac-based x-ray fel facility in shanghai.
\newblock In {\em Proc. of FEL2017, Santa Fe, NM, USA}, pages 20--25, 2017.

\bibitem{serafini2019marix}
L~Serafini et~al.
\newblock Marix, an advanced mhz-class repetition rate x-ray source for linear
  regime time-resolved spectroscopy and photon scattering.
\newblock {\em Nucl. Instrum. Meth. A}, 930:167--172, 2019.

\bibitem{kondratenko1980generation}
AM~Kondratenko and EL~Saldin.
\newblock Generation of coherent radiation by a relativistic electron beam in
  an ondulator.
\newblock {\em Part. Accel.}, 10(3-4):207--216, 1980.

\bibitem{bonifacio1984collective}
R.~Bonifacio, C.~Pellegrini, and L.~M. Narducci.
\newblock Collective instabilities and high-gain regime in a free electron
  laser.
\newblock {\em Opt. Commun.}, 50:373--378, 1984.

\bibitem{feldhaus1997possible}
J~Feldhaus, EL~Saldin, JR~Schneider, EA~Schneidmiller, and MV~Yurkov.
\newblock Possible application of x-ray optical elements for reducing the
  spectral bandwidth of an x-ray sase fel.
\newblock {\em Opt. Commun.}, 140(4-6):341--352, 1997.

\bibitem{saldin2001x}
EL~Saldin, EA~Schneidmiller, Yu~V Shvyd'ko, and MV~Yurkov.
\newblock X-ray fel with a mev bandwidth.
\newblock {\em Nucl. Instrum. Meth. A}, 475(1-3):357--362, 2001.

\bibitem{geloni2011novel}
Gianluca Geloni, Vitali Kocharyan, and Evgeni Saldin.
\newblock A novel self-seeding scheme for hard x-ray fels.
\newblock {\em J. Mod. Opt.}, 58(16):1391--1403, 2011.

\bibitem{lambert2008injection}
G~Lambert, T~Hara, D~Garzella, T~Tanikawa, M~Labat, B~Carre, H~Kitamura,
  T~Shintake, M~Bougeard, S~Inoue, et~al.
\newblock Injection of harmonics generated in gas in a free-electron laser
  providing intense and coherent extreme-ultraviolet light.
\newblock {\em Nat. phys.}, 4(4):296--300, 2008.

\bibitem{togashi2011extreme}
Tadashi Togashi, Eiji~J Takahashi, Katsumi Midorikawa, Makoto Aoyama, Koichi
  Yamakawa, Takahiro Sato, Atsushi Iwasaki, Shigeki Owada, Tomoya Okino, Kaoru
  Yamanouchi, et~al.
\newblock Extreme ultraviolet free electron laser seeded with high-order
  harmonic of ti: sapphire laser.
\newblock {\em Opt. Express}, 19(1):317--324, 2011.

\bibitem{mcpherson1987studies}
A~McPherson, G~Gibson, H~Jara, U~Johann, Ting~S Luk, IA~McIntyre, Keith Boyer,
  and Charles~K Rhodes.
\newblock Studies of multiphoton production of vacuum-ultraviolet radiation in
  the rare gases.
\newblock {\em JOSA B}, 4(4):595--601, 1987.

\bibitem{ferray1988multiple}
M~Ferray, Anne L'Huillier, XF~Li, LA~Lompre, G~Mainfray, and C~Manus.
\newblock Multiple-harmonic conversion of 1064 nm radiation in rare gases.
\newblock {\em J. Phys. B: At., Mol. Opt. Phys.}, 21(3):L31, 1988.

\bibitem{seres2005source}
Jozsef Seres, E~Seres, Aart~J Verhoef, G~Tempea, Ch~Streli, P~Wobrauschek,
  V~Yakovlev, Armin Scrinzi, Ch~Spielmann, and Ferenc Krausz.
\newblock Source of coherent kiloelectronvolt x-rays.
\newblock {\em Nature}, 433(7026):596--596, 2005.

\bibitem{labaye2017extreme}
Fran{\c{c}}ois Labaye et~al.
\newblock Extreme ultraviolet light source at a megahertz repetition rate based
  on high-harmonic generation inside a mode-locked thin-disk laser oscillator.
\newblock {\em Opt. Lett.}, 42(24):5170--5173, 2017.

\bibitem{yu1991generation}
Li~Hua Yu.
\newblock Generation of intense uv radiation by subharmonically seeded
  single-pass free-electron lasers.
\newblock {\em Phys. Rev. A}, 44(8):5178, 1991.

\bibitem{yu2000high}
L-H Yu et~al.
\newblock High-gain harmonic-generation free-electron laser.
\newblock {\em Science}, 289(5481):932--934, 2000.

\bibitem{allaria2012highly}
E~Allaria et~al.
\newblock Highly coherent and stable pulses from the fermi seeded free-electron
  laser in the extreme ultraviolet.
\newblock {\em Nat. Photon.}, 6(10):699--704, 2012.

\bibitem{stupakov2009using}
Gennady Stupakov.
\newblock Using the beam-echo effect for generation of short-wavelength
  radiation.
\newblock {\em Phys. Rev. Lett.}, 102(7):074801, 2009.

\bibitem{xiang2009echo}
Dao Xiang and Gennady Stupakov.
\newblock Echo-enabled harmonic generation free electron laser.
\newblock {\em Phys. Rev. ST Accel. Beams}, 12(3):030702, 2009.

\bibitem{ribivc2019coherent}
Primo{\v{z}}~Rebernik Ribi{\v{c}} et~al.
\newblock Coherent soft x-ray pulses from an echo-enabled harmonic generation
  free-electron laser.
\newblock {\em Nat. Photon.}, 13(8):555--561, 2019.

\bibitem{penco2020enhanced}
Giuseppe Penco, G~Perosa, E~Allaria, S~Di~Mitri, E~Ferrari, L~Giannessi,
  S~Spampinati, C~Spezzani, and M~Veronese.
\newblock Enhanced seeded free electron laser performance with a “cold”
  electron beam.
\newblock {\em Phys. Rev. Accel. Beams}, 23(12):120704, 2020.

\bibitem{feng2019coherent}
Chao Feng et~al.
\newblock Coherent extreme ultraviolet free-electron laser with echo-enabled
  harmonic generation.
\newblock {\em Phys. Rev. Accel. Beams}, 22(5):050703, 2019.

\bibitem{kinoshita2014development}
Hiroo Kinoshita, Tetsuo Harada, Yutaka Nagata, Takeo Watanabe, and Katsumi
  Midorikawa.
\newblock Development of euv mask inspection system using high-order harmonic
  generation with a femtosecond laser.
\newblock {\em Jpn. J. Appl. Phys.}, 53(8):086701, 2014.

\bibitem{nagata2019wavelength}
Yutaka Nagata, Tetsuo Harada, Takeo Watanabe, Hiroo Kinoshita, and Katsumi
  Midorikawa.
\newblock At wavelength coherent scatterometry microscope using high-order
  harmonics for euv mask inspection.
\newblock {\em Int. J. Extreme Manuf.}, 1(3):032001, 2019.

\bibitem{erik}
Erik~R. Hosler, Obert R.~Wood II, and William~A. Barletta.
\newblock {Free-electron laser emission architecture impact on EUV
  lithography}.
\newblock In Eric~M. Panning, editor, {\em Extreme Ultraviolet Lithogr. VIII},
  volume 10143, pages 338 -- 349. Int. Soc. Opt. Photon., SPIE, 2017.

\bibitem{mochi2019lensless}
Iacopo Mochi and Yasin Ekinci.
\newblock Lensless euv lithography and imaging.
\newblock {\em Synchrotron Radiat. News}, 32(4):22--27, 2019.

\bibitem{kudilatt2020quantum}
Hasna Kudilatt, Bo~Hou, and Mark~E Welland.
\newblock Quantum dots microstructural metrology: From time-resolved
  spectroscopy to spatially resolved electron microscopy.
\newblock {\em Part. Part. Syst. Charact.}, 37(12):2000192, 2020.

\bibitem{petrillo2019high}
Vittoria Petrillo et~al.
\newblock High repetition rate and coherent free-electron laser in the x-rays
  range tailored for linear spectroscopy.
\newblock {\em Instruments}, 3(3):47, 2019.

\bibitem{pedersoli2013mesoscale}
E~Pedersoli et~al.
\newblock Mesoscale morphology of airborne core--shell nanoparticle clusters:
  X-ray laser coherent diffraction imaging.
\newblock {\em J. Phys. B: At., Mol. Opt. Phys.}, 46(16):164033, 2013.

\bibitem{capotondi2015multipurpose}
Flavio Capotondi et~al.
\newblock Multipurpose end-station for coherent diffraction imaging and
  scattering at fermi@ elettra free-electron laser facility.
\newblock {\em J. Synchrotron Radiat.}, 22(3):544--552, 2015.

\bibitem{helfenstein2017coherent}
Patrick Helfenstein, Iacopo Mochi, Rajendran Rajeev, Sara Fernandez, and Yasin
  Ekinci.
\newblock Coherent diffractive imaging methods for semiconductor manufacturing.
\newblock {\em Adv. Opt. Technol.}, 6(6):439--448, 2017.

\bibitem{attwood2000soft}
David Attwood.
\newblock {\em Soft x-rays and extreme ultraviolet radiation: principles and
  applications}.
\newblock Cambridge university press, 2000.

\bibitem{bencivenga2015four}
F~Bencivenga et~al.
\newblock Four-wave mixing experiments with extreme ultraviolet transient
  gratings.
\newblock {\em Nature}, 520(7546):205--208, 2015.

\bibitem{chang2020ultrafast}
Guoqing Chang and Zhiyi Wei.
\newblock Ultrafast fiber lasers: an expanding versatile toolbox.
\newblock {\em Iscience}, 23(5):101101, 2020.

\bibitem{riedel2013long}
R~Riedel, M~Schulz, MJ~Prandolini, A~Hage, H~H{\"o}ppner, T~Gottschall,
  J~Limpert, M~Drescher, and F~Tavella.
\newblock Long-term stabilization of high power optical parametric
  chirped-pulse amplifiers.
\newblock {\em Opt. Express}, 21(23):28987--28999, 2013.

\bibitem{prinz2015cep}
Stephan Prinz et~al.
\newblock Cep-stable, sub-6 fs, 300-khz opcpa system with more than 15 w of
  average power.
\newblock {\em Opt. Express}, 23(2):1388--1394, 2015.

\bibitem{puppin2015500}
Michele Puppin, Yunpei Deng, Oliver Prochnow, Jan Ahrens, Thomas Binhammer, Uwe
  Morgner, Marcel Krenz, Martin Wolf, and Ralph Ernstorfer.
\newblock 500 khz opcpa delivering tunable sub-20 fs pulses with 15 w average
  power based on an all-ytterbium laser.
\newblock {\em Opt. Express}, 23(2):1491--1497, 2015.

\bibitem{hoppner2015optical}
H~H{\"o}ppner, A~Hage, T~Tanikawa, M~Schulz, R~Riedel, U~Teubner,
  MJ~Prandolini, B~Faatz, and F~Tavella.
\newblock An optical parametric chirped-pulse amplifier for seeding high
  repetition rate free-electron lasers.
\newblock {\em New J. Phys.}, 17(5):053020, 2015.

\bibitem{mecseki2019high}
Katalin Mecseki, Matthew~KR Windeler, Alan Miahnahri, Joseph~S Robinson,
  James~M Fraser, Alan~R Fry, and Franz Tavella.
\newblock High average power 88 w opcpa system for high-repetition-rate
  experiments at the lcls x-ray free-electron laser.
\newblock {\em Opt. Lett.}, 44(5):1257--1260, 2019.

\bibitem{reinsch2012radiator}
M~Reinsch, G~Penn, P~Gandhi, and J~Wurtele.
\newblock The radiator-first hghg multi-mhz x-ray fel concept.
\newblock In {\em Proc. of FEL2012, Nara, Japan}, pages 273--276, 2012.

\bibitem{li2018high}
Kai Li, Jiawei Yan, Chao Feng, Meng Zhang, and Haixiao Deng.
\newblock High brightness fully coherent x-ray amplifier seeded by a
  free-electron laser oscillator.
\newblock {\em Phys. Rev. Accel. Beams}, 21(4):040702, 2018.

\bibitem{petrillo2020coherent}
V~Petrillo et~al.
\newblock Coherent, high repetition rate tender x-ray free-electron laser
  seeded by an extreme ultra-violet free-electron laser oscillator.
\newblock {\em New J. Phys.}, 22(7):073058, 2020.

\bibitem{ackermann2020novel}
Sven Ackermann et~al.
\newblock Novel method for the generation of stable radiation from
  free-electron lasers at high repetition rates.
\newblock {\em Phys. Rev. Accel. Beams}, 23(7):071302, 2020.

\bibitem{paraskaki2021optimization}
Georgia Paraskaki, Vanessa Grattoni, Tino Lang, Johann Zemella, Bart Faatz, and
  Wolfgang Hillert.
\newblock Optimization and stability of a high-gain harmonic generation seeded
  oscillator amplifier.
\newblock {\em Phys. Rev. Accel. Beams}, 24(3):034801, 2021.

\bibitem{mirian2021high}
NS~Mirian, M~Opromolla, G~Rossi, L~Serafini, and V~Petrillo.
\newblock High-repetition rate and coherent free-electron laser in the tender x
  rays based on the echo-enabled harmonic generation of an ultraviolet
  oscillator pulse.
\newblock {\em Phys. Rev. Accel. Beams}, 24(5):050702, 2021.

\bibitem{feng2017storage}
Chao Feng and Zhentang Zhao.
\newblock A storage ring based free-electron laser for generating ultrashort
  coherent euv and x-ray radiation.
\newblock {\em Sci. Rep.}, 7(1):1--7, 2017.

\bibitem{wang2019angular}
Xiaofan Wang, Chao Feng, Tao Liu, Zhen Zhang, C-Y Tsai, Juhao Wu, Chuan Yang,
  and Zhentang Zhao.
\newblock Angular dispersion enhanced prebunch for seeding ultrashort and
  coherent euv and soft x-ray free-electron laser in storage rings.
\newblock {\em J. Synchrotron Radiat.}, 26(3):677--684, 2019.

\bibitem{dunning2011design}
DJ~Dunning, NR~Thompson, and BWJ McNeil.
\newblock Design study of an hhg-seeded harmonic cascade free-electron laser.
\newblock {\em J. Mod. Opt.}, 58(16):1362--1373, 2011.

\bibitem{yan2021self}
Jiawei Yan et~al.
\newblock Self-amplification of coherent energy modulation in seeded
  free-electron lasers.
\newblock {\em Phys. Rev. Lett.}, 126(8):084801, 2021.

\bibitem{paraskaki2021high}
Georgia Paraskaki, Enrico Allaria, Evgeny Schneidmiller, and Wolfgang Hillert.
\newblock High repetition rate seeded free electron laser with an optical
  klystron in high-gain harmonic generation.
\newblock {\em Phys. Rev. Accel. Beams}, 24(12):120701, 2021.

\bibitem{reiche1999genesis}
Sven Reiche.
\newblock Genesis 1.3: a fully 3d time-dependent fel simulation code.
\newblock {\em Nucl. Instrum. Meth. A}, 429(1-3):243--248, 1999.

\bibitem{leontowich2021lower}
Adam~FG Leontowich, Ariel Gomez, Beatriz Diaz~Moreno, David Muir, Denis
  Spasyuk, Graham King, Joel~W Reid, C-Y Kim, and Stefan Kycia.
\newblock The lower energy diffraction and scattering side-bounce beamline for
  materials science at the canadian light source.
\newblock {\em J. Synchrotron Radiat.}, 28(3):961--969, 2021.

\bibitem{feng2012study}
C~Feng, D~Wang, and ZT~Zhao.
\newblock Study of the energy chirp effects on seeded fel schemes at sduv-fel.
\newblock In {\em Proc. of IPAC12, New Orleans, LA, USA}, pages 1724--1726,
  2012.

\bibitem{hemsing2014highly}
E~Hemsing, M~Dunning, C~Hast, TO~Raubenheimer, S~Weathersby, and D~Xiang.
\newblock Highly coherent vacuum ultraviolet radiation at the 15th harmonic
  with echo-enabled harmonic generation technique.
\newblock {\em Phys. Rev. ST Accel. Beams}, 17(7):070702, 2014.

\bibitem{paraskaki2019impact}
Georgia Paraskaki, Sven Ackermann, Bart Faatz, Vanessa Grattoni, Christoph
  Lechner, Johann Zemella, and Germany~W Hillert.
\newblock Impact of electron beam energy chirp on seeded fels.
\newblock In {\em Proc. of FEL19, Hamburg, Germany}, pages 238--241, 2019.

\bibitem{feng2016single}
Chao Feng, Dazhang Huang, Haixiao Deng, Jianhui Chen, Dao Xiang, Bo~Liu, Dong
  Wang, and Zhentang Zhao.
\newblock A single stage eehg at sxfel for narrow-bandwidth soft x-ray
  generation.
\newblock {\em Sci. Bull.}, 61(15):1202--1212, 2016.

\end{thebibliography}

\end{document}